\documentclass[%
 aip,
 amsmath,amssymb,
 reprint,%
]{revtex4-1}

\usepackage{graphicx}% Include figure files
\usepackage{dcolumn}% Align table columns on decimal point
\usepackage{bm}% bold math
\usepackage{tabularx}
\usepackage[utf8]{inputenc}
\usepackage[T1]{fontenc}
\usepackage{mathptmx}
\usepackage{etoolbox}

\makeatletter
\def\@email#1#2{%
 \endgroup
 \patchcmd{\titleblock@produce}
  {\frontmatter@RRAPformat}
  {\frontmatter@RRAPformat{\produce@RRAP{*#1\href{mailto:#2}{#2}}}\frontmatter@RRAPformat}
  {}{}
}%
\makeatother
\begin{document}

\preprint{AIP/123-QED}

\title{A Physical Unclonable Function Based on Variations of Write Times in STT-MRAM due to Manufacturing Defects}
% Force line breaks with \\

\author{Jacob Huber}
 \affiliation {Department of Electrical and Computer Engineering, Virginia Commonwealth University, Richmond, VA 23284, USA}%Lines break automatically or can be forced with \\
\author{Supriyo Bandyopadhyay*}
\email{sbandy@vcu.edu}
\affiliation {Department of Electrical and Computer Engineering, Virginia Commonwealth University, Richmond, VA 23284, USA}%Lines break automatically or can be forced with \\

\date{\today}% It is always \today, today,

\begin{abstract}
A physical unclonable function (PUF) utilizes the unclonable random variations in a device's responses to a set of inputs  to produce a unique ``biometric'' that can be used for authentication. The variations are caused by unpredictable, unclonable and random manufacturing defects. Here, we show that the switching time of a magnetic tunnel junction injected with a spin-polarized current generating spin transfer torque is sensitive to the nature of structural defects introduced during manufacturing and hence can be the basis of a PUF. We use micromagnetic simulations to study the switching times under a constant current excitation for six different (commonly encountered) defect morphologies in spin-transfer-torque magnetic random access memory (STT-MRAM) to establish the viability of a PUF.  
\end{abstract}

\maketitle

%\section{\label{sec:level1}Introduction:\protect\\  \lowercase{} \textbackslash\textbackslash}
\section{Introduction}

Electronic devices often require authentication for trust. An efficient strategy for this is to enable authentication via the physical characteristics of a device that are sensitive to manufacturing defects \cite{devadas}. These manufacturing defects are unpredictable and impossible to anticipate, duplicate, or hack. The defects determine the responses (outputs) of the device to a set of challenges (inputs) so this ``challenge-response'' set becomes a signature of the device's defect morphology. That makes it a fingerprint or biometric of the device, which can be used for authentication.

Magnetic tunnel junctions (MTJs), which are the mainstay of STT-MRAM, have recently emerged as a popular rendition of PUFs \cite{huber,chen,cao,kumar,das,zhang,marukame,vatajelu}. An MTJ is a three-layered structure with the two outer layers ferromagnetic and the middle layer a non-magnetic insulator through which spin-dependent tunneling of electrons takes place. The MTJ resistance is high when the magnetizations of the two ferromagnetic layers are antiparallel and low when they are parallel. One ferromagnetic layer (the ``hard'' layer) has a fixed magnetization while the other (the ``soft'' layer) has bistable magnetization which can point in one of two (mutually antiparallel) directions. Flipping the magnetization of the soft layer from one stable direction to the other results in switching the resistance of the MTJ from high to low, or vice versa. 

There are many ways to flip the magnetization of the soft layer, but the one commonly employed is referred to as ``spin-transfer-torque'' (STT). This involves passing a spin polarized current of fixed amplitude through the MTJ for a certain duration of time (a current pulse). The electrons in the current pulse transfer their angular momenta to the resident electrons in the  soft layer and align the latter's magnetization along one of the two stable directions depending on the current polarity. This will will set the MTJ resistance to either the high or the low value, thus accomplishing the ``switching'' action. 

The probability of switching of a {\it defect-free} pristine MTJ due to STT will depend on the amplitude of the current pulse as well as the pulse width. For a fixed switching probability, if we fix the amplitude of the current, then the pulse width will also be fixed. However, that changes in the presence of random defects. For a fixed current amplitude,  the pulse duration or width required to achieve a given switching probability will depend on the concentration and nature of the defects. 

Since the defects are incurred during manufacturing, their morphology and concentration are unpredictable and unclonable. That will make the pulse width required to achieve a given switching probability with a given current amplitude unpredictable and unclonable. 

Let us say that we subject an MTJ to a current pulse of amplitude $I$ and width of $T_1$ repeatedly, and then check how many times it has switched in every trial to obtain the probability of switching. We then repeat this exercise for a pulse width of $T_2$ and so on. We can then produce a table of pulse width versus switching probability for a fixed pulse amplitude. This table, which will be the challenge-response set for our MTJ, will be {\it different for different MTJs} with different defect morphologies and  concentrations, and hence will make a unique  ``fingerprint'' for that MTJ. This is the basis of a PUF. This same principle has been used in \cite{kumar,rose}.

\section{PUFs based on pulse width dependence on defect morphology}

\begin{figure*}[hbt!]
\includegraphics[width=7in]{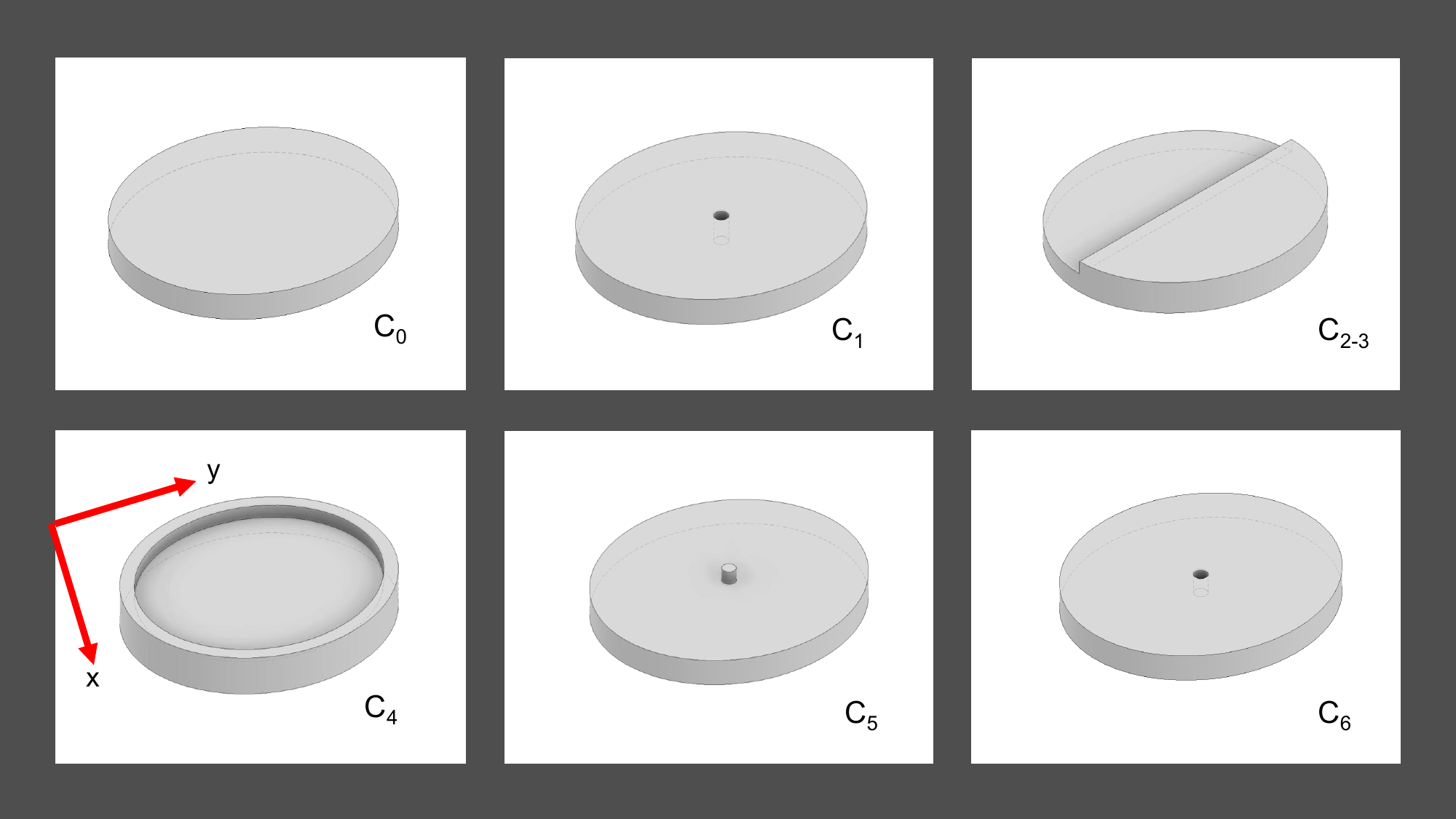}
\caption{Six different defect morphologies in an elliptical cobalt soft layer of major axis 100 nm, minor axis 90 nm, and average thickness 3 nm. $C_0$ is a defect-free pristine elliptical soft layer; $C_1$ has a 5-nm diameter hole in the center that is 2 nm deep; $C_{2-3}$ has one-half thicker than the other, with the thicker half 4 nm and the thinner half 2 nm for a step size of 2 nm; $C_4$ has a 10-nm wide rim that rises 1 nm above the surface; $C_5$ has a 5-nm diameter rivet that rises 1 nm above the surface; finally $C_6$ has through-hole with a diameter of 5 nm. Reproduced from [2] with a CC-BY 4.0 license.}
\label{fig:defects}
\end{figure*}

We evaluate the switching probability of a STT-MRAM MTJ with a given defect morphology as a function of current pulse width for a fixed current amplitude to establish the viability of a PUF. The soft layer is assumed to be made of cobalt and shaped like an elliptical disk of major axis 100 nm, minor axis 90 nm, and average thickness 3 nm. The soft layer has manufacturing defects and we consider six different defect morphologies associated with voids, random thickness variations, etc. These defects are not chosen arbitrarily but conform to defects actually observed with atomic force microscopy in fabricated samples \cite{winters}. Their models are shown in Fig. \ref{fig:defects}.  We will consider one pristine (defect-free) soft layer and five defective soft layers, each having a different type of defect. 

To calculate the switching probability at room temperature as a function of current pulse width for a fixed current amplitude for each defect morphology, we use the micromagnetic simulator MuMax3  \cite{mumax} with a thermal noise term included. 
The magnetization of the soft layer is initially aligned along one of the two stable directions along the major axis (or easy axis), say the --y-direction as shown in Fig. \ref{fig:defects}, with an external magnetic field. Next, a 3 mA spin polarized current  with a majority of the spins polarized along the +y-direction, is injected perpendicular to the surface.  We simulate the time evolution of the magnetization which is governed by the Landau-Lifshitz-Gilbert-Langevin equation (within the MuMax3 simulator) and track the temporal evolution of the magnetization for a fixed duration (current pulse width) in the presence of thermal noise. Flipping takes place when the magnetization aligns close to the +y-direction and the normalized y-component of the magnetization exceeds 0.9, regardless of what the x- and z-components are.

For the purpose of simulation, we map the different defect morphologies on to the MuMax3 software package using a grid divided into 32 $\times$ 32 $\times$ 4 cells, each cell being  2.8 nm $\times$ 3.1 nm $\times$ 1 nm. This gives a total volume of 90 nm $\times$ 100 nm $\times$ 4 nm. The ambient temperature is assumed to be 300 K. An effective magnetic field due to thermal noise at 300 K is introduced as \cite{brown}
\begin{equation*}
    h_i^{noise}(t) = \sqrt{{{2 \alpha kT}\over{\gamma \left (1 + \alpha^2 \right ) \mu_0 M_s \Omega \Delta t}}}G_{(0,1)}^i (t),
\end{equation*}
 where $\alpha$ is the Gilbert damping constant, $\gamma$ is the universal gyromagnetic factor, $G_{(0,1)}^i (t)$ ($i = x, y, z$) are three uncorrelated  Gaussians of zero mean and unit standard deviation, $\Omega$ is the nanomagnet volume, $k$ is the Boltzmann constant, $T$ is the absolute temperature, and $\Delta t$ is the  time step of the simulation (0.09 ps). We use cobalt's material parameters. The current is spin polarized and we assume the spin polarization percentage to be 30\%. 

\begin{figure*}[hbt!]
\includegraphics[width=7in]{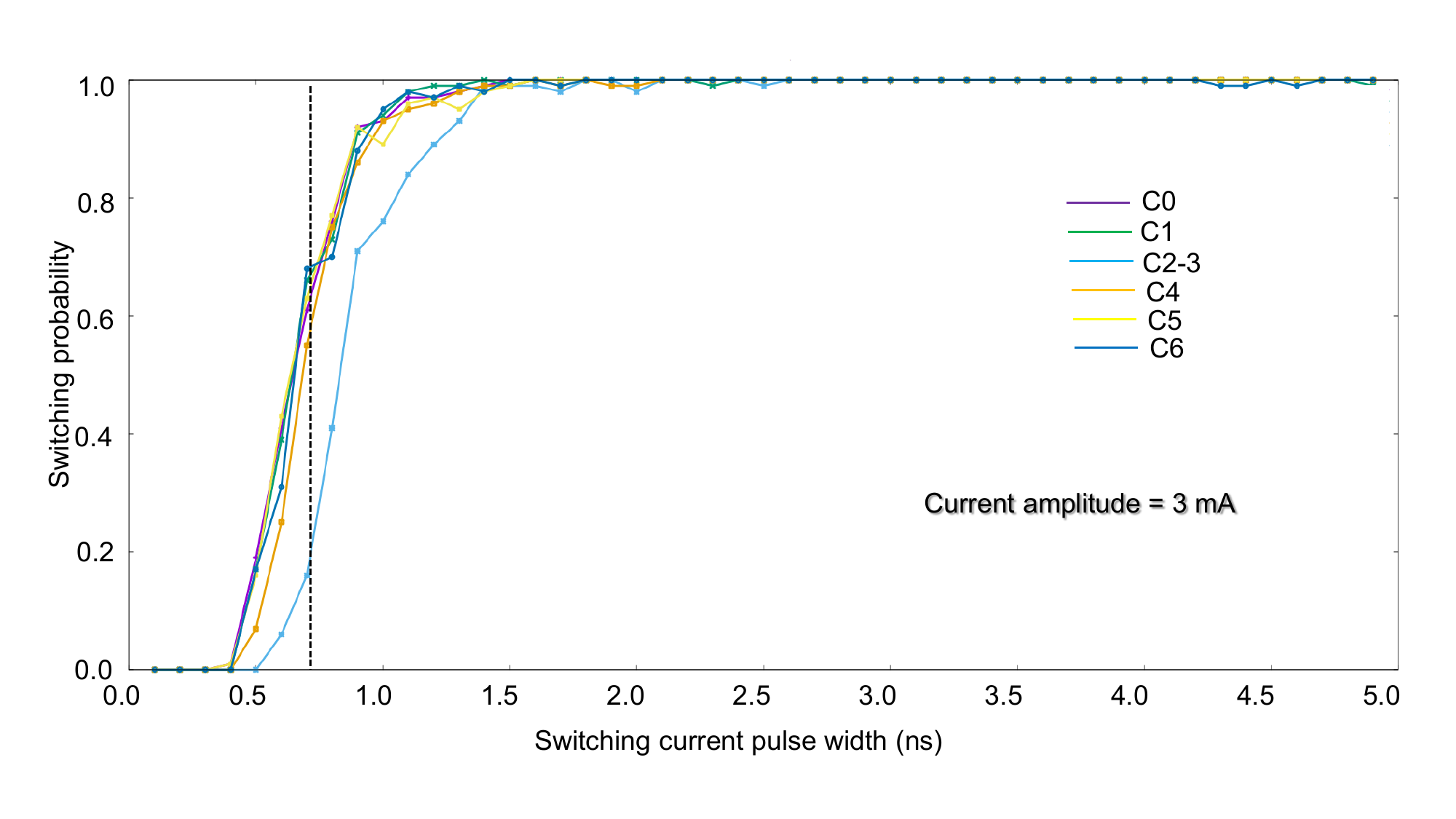}
\caption{Switching probability for the six different defect morphologies as a function of current pulse width for a current amplitude of 3 mA at a temperature of 300 K. The dashed line shows the probabilities for a pulse width of 0.75 ns.}
\label{fig:plot}
\end{figure*}
 
 The thermal noise makes the switching trajectories random. Hence we generate 100 trajectories and the probability of switching is defined as the fraction of the trajectories that at the end of the pulse duration show a y-component of normalized magnetization greater than or equal to 0.9.  We calculate these probabilities for different current pulse durations and for a fixed pulse amplitude of 3 mA. They are calculated for the six different defect morphologies shown in Fig. \ref{fig:defects} and the resulting plot of switching probability versus pulse duration is shown in Fig. \ref{fig:plot}.

From Fig. \ref{fig:plot}, we can construct a table of switching probability at any pulse width and amplitude for the six different defect morphologies. For example, given a pulse width of 0.75 ns and pulse amplitude of 3 mA, the table will read as shown in Table I.

\begin{table}[!h]
\centering
    \caption{\bf Switching probabilities for different defect morphologies at 300 K temperature for a 3 mA current amplitude and 0.75 ns current pulse duration}
    \vspace{3mm}
    \begin{tabular}{|c|c|}
    \hline
    Defect & Switching probability (\%)   \\
    \hline
    C$_0$ & 63 \\
    C$_1$ & 69 \\
    C$_{2-3}$ & 20  \\
    C$_4$ & 57  \\
    C$_5$ & 63 \\
    C$_6$ & 69 \\
    \hline
    \end{tabular}
    \label{tab:table1}
\end{table}

We can then proceed to construct a similar table for a different pulse duration at the same pulse amplitude. These tables form the {\it challenge-response} set. 

\subsection{PUF Implementation}

Consider a sample with three MTJs. A magnetic field or a high current of the correct polarity will reset all three MTJs to the high resistance state. Each is then subjected to a current pulse of the opposite polarity in an attempt to switch them to the low resistance state. This current pulse's amplitude is 3 mA and the width is 0.75 ns. We will assume that MTJ-1 has a defect of type $C_1$, MTJ-2 has a defect 
of type $C_4$  and MTJ-3 has a defect of type $C_{2-3}$. The probabilities that MTJ-1 and MTJ-2 will switch are relatively high (69\% and 57\%, respectively)  while the probability of MTJ-3 switching is low (20\%). Following ref. [\onlinecite{kumar}], we will assign the bit 1 to high probability  and bit 0 to low probability. The switching ``response'' to this current pulse ``challenge'' can therefore be coded as 110. 

Let us take another sample where the defect morphologies of MTJ-2 and MTJ-3 are switched. In that case, the response bit stream will be 101. 

We can generate the response bit streams for all possible combinations of defect morphologies in a 3-MTJ unit (each MTJ has a different defect $\in (C_1, C_4, C_{2-3})$ following the convention that the first bit in the response bit stream corresponds to the switching probability (high or low)  of the first MTJ, the second to that of the second MTJ, and the third to that of the third MTJ. Using this convention, we can generate the response bit streams for six different 3-MTJ units each having a different defect distribution $\in (C_1, C_4, C_{2-3})$) (no defect type repeated in two or more MTJs). This is shown in Table II. 

\begin{table}[!h]
\centering
    \caption{\bf Response bit streams under a current pulse of amplitude 3 mA and width 0.75 ns for six different 3-MTJ units with different distribution of the defect morphologies $\in (C_1, C_4, C_{2-3})$) at 300 K temperature}
    \vspace{3mm}
    \begin{tabular}{|c|c|}
    \hline
    Defect distribution & Response bit stream  \\
    \hline
    Unit 1: C$_1$ ~C$_{4}$ ~C$_{2-3}$ & 110 \\
    Unit 2: C$_1$ ~C$_{2-3}$ ~C$_{4}$ & 101 \\
    Unit 3: C$_{4}$ ~C$_1$ ~C$_{2-3}$ & 110 \\
    Unit 4: C$_{4}$ ~C$_{2-3}$ ~C$_1$ & 101 \\
    Unit 5: C$_{2-3}$ ~C$_{4}$ ~C$_1$ & 011 \\
    Unit 6: C$_{2-3}$ ~C$_1$ ~C$_{4}$ & 011 \\
    \hline
    \end{tabular}
    \label{tab:table2}
\end{table}

It is easy to see that if the pulse width is reduced to 0.7 ns, then both Tables \ref{tab:table1} and \ref{tab:table2} will change:

\begin{table}[!h]
\centering
    \caption{\bf Switching probabilities for different defect morphologies at 300 K temperature for 3 mA current amplitude and 0.7 ns current pulse duration}
    \vspace{3mm}
    \begin{tabular}{|c|c|}
    \hline
    Defect & Switching probability (\%)   \\
    \hline
    C$_0$ & 55 \\
    C$_1$ & 55 \\
    C$_{2-3}$ & 15  \\
    C$_4$ & 45  \\
    C$_5$ & 55 \\
    C$_6$ & 56 \\
    \hline
    \end{tabular}
\end{table}

\begin{table}[!h]
\centering
    \caption{\bf Response bit streams under a current pulse of amplitude 3 mA and width 0.7 ns for six different 3-MTJ units with different distribution of the defect morphologies $\in (C_1, C_4, C_{2-3})$) at 300 K temperature}
    \vspace{3mm}
    \begin{tabular}{|c|c|}
    \hline
    Defect distribution & Response bit stream  \\
    \hline
    Unit 1: C$_1$ ~C$_{4}$ ~C$_{2-3}$ & 100 \\
    Unit 2: C$_1$ ~C$_{2-3}$ ~C$_{4}$ & 100 \\
    Unit 3: C$_{4}$ ~C$_1$ ~C$_{2-3}$ & 010 \\
    Unit 4: C$_{4}$ ~C$_{2-3}$ ~C$_1$ & 001 \\
    Unit 5: C$_{2-3}$ ~C$_{4}$ ~C$_1$ & 001 \\
    Unit 6: C$_{2-3}$ ~C$_1$ ~C$_{4}$ & 010 \\
    \hline
    \end{tabular}
\end{table}

Thus, by using different challenges and the corresponding responses, we can generate a challenge-response table which will depend on the defect morphologies and hence will be a fingerprint of the specific unit under consideration. Since the defect morphologies are unpredictable, unknown, and unclonable, the challenge-response table is also unpredictable, unknown, and unclonable. One will have to measure the response to each challenge and establish the unique challenge-response characteristic of any unit which then becomes a biometric of that unit and enables a PUF.

\subsection{Inter-Hamming distance}

The inter-Hamming distance (IHD) is defined as the number of positions where the bits in the response bit stream are different for two different units divided by the number of bits, averaged over all possible pairs. We can see from Table 2, that the Hamming distance between units 1 and 2 is 2/3, between 1 and 3 is 0, between 1 and 4 is 2/3, between 1 and 5 is 2/3, and between 1 and 6 is also 2/3. 

Similarly, between units 2 and 3, it is 2/3, between 2 and 4 it is 0, between 2 and 5 it is 2/3, and between 2 and 6 it is 2/3. 

\begin{figure*}[!hbt]
\includegraphics[width=7in]{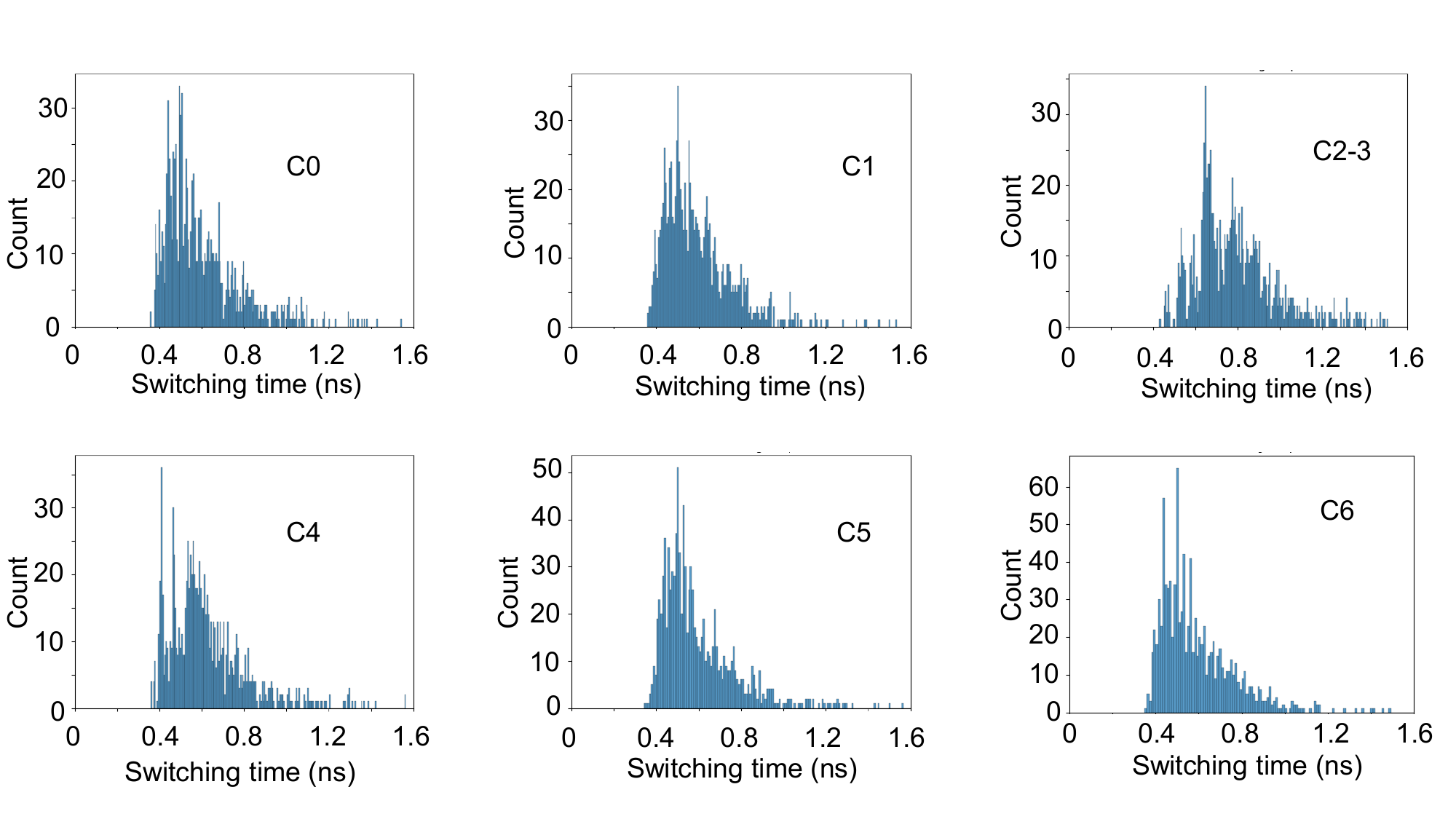}
\caption{Switching time distribution for the six different defect morphologies for a current pulse amplitude of 3 mA at a temperature of 300 K. }
\label{fig:dist}
\end{figure*}

We can go on with the other combinations and then average over all combinations to find that the IHD is 8/15 or 0.533. This is only true for the challenge current of 3 mA amplitude and 0.75 ns pulse width, and will change for a different challenge. The ideal IHD for a PUF is 0.5~ \cite{chen}.

In this example, we arbitrarily assumed that no two MTJs in a unit have the same defect morphology. In reality, this restriction should not exist and the defect morphology can be repeated in two or even all three MTJs. In that case, we will have 3$^3$ = 27 possible combinations (27 different units) and then the IHD will have to be computed by comparing the response bit streams of $(n/2)(n+1)$ [$n$ = 26], i.e. 351 pairs. This process leads to an IHD = 0.479 for the challenge current of 3 mA and 0.75 ns pulse width. A different challenge current will, of course, produce a different IHD.

The IHDs we calculate are very similar to those reported in the experiments of ref. [\onlinecite{chen}] and [\onlinecite{cao}], which reported values of 0.498 and 0.53, respectively.

\subsection{Other Implementations Based on Dependence of Switching Time Probability Distribution on Defect Morphology}

More sophisticated PUFs can be implemented by measuring ``distributions'' rather than single quantities. One possibility is to determine the switching time distribution for a fixed current amplitude and for a fixed switching probability. These distributions are sensitive to defect morphology and hence can be the basis of a strong PUF. In Fig. \ref{fig:dist} we show the switching time distribution for a current of amplitude 3 mA for the six different defect morphologies across 1000 trials for each morphology. Once again, ``switching'' is defined as
the y-component of normalized magnetization becoming greater than or equal to 0.9 when the initial y-component is -1.

\section{Conclusion}

We have shown that the sensitivity of the switching time of an MTJ (or STT-MRAM) to defect morphology can be exploited to implement physical unclonable functions.  The inter-Hamming distances are close to ideal values, which makes these PUFs reliable for authentication.

\section*{Acknowledgments}
J. Huber was supported by a Research Experience for Undergraduates (REU) grant [REU Site in Magnetics] from the US National Science Foundation (grant number DMR-2349694). High Performance Computing resources provided by the High Performance Research Computing (HPRC) core facility at Virginia Commonwealth University (https://hprc.vcu.edu) were used for conducting the research reported in this work.

\section*{Author contributions}
J. H. carried out all the simulations. S. B. 
conceived the idea and supervised the project. All authors contributed to writing the paper.

\section*{Conflict of Interests}
The authors declare no conflict of interests.

\vspace{0.2in}

\section*{Data Availability Statement} 

All data generated are already available in the paper.


\begin{thebibliography}{10}

\bibitem{devadas}
C. Herder, M-D Yu, K. Farinaz and S.Devadas, ``Physical Unclonable Functions
and Applications: A Tutorial'', {\it Proc. IEEE}, {\bf 102}, 1126 (2014).

\bibitem{huber}
J. Huber, R. Rahman and S. Bandyopadhyay, ``Sensitivity of the Threshold Current for Switching of a Magnetic
Tunnel Junction to Fabrication Defects and its Application in
Physical Unclonable Functions'', {\it Appl. Sci.} {\bf 15}, 9548 , (2025).

\bibitem{chen}
H. Chen, M. Song, Z. Guo, R. Li, Q.  Zou, S.  Luo, S. Zhang, Q. Luo, J. Hong and L. You, ``Highly Secure Physically Unclonable Cryptographic Primitives Based
on Interfacial Magnetic Anisotropy'', {\it  Nano Lett.}, {\bf 18}, 7211 (2018).

\bibitem{cao}
Z. Cao, S. Zhang, J. Zhang, N Xu, R. Li, Z. Guo, J. Yun, M. Song, Q. Zou, L. Xi and O. Lee, ``Reconfigurable Physical Unclonable Function
Based on Spin-Orbit Torque Induced Chiral
Domain Wall Motion'' {\it IEEE Elec. Dev. Lett..},  {\bf 42} 597 (2021).

\bibitem{kumar}
A. Kumar, S. Sahay and M. Suri,  ``Switching-Time Dependent PUF Using
STT-MRAM'', 31th International Conference on VLSI Design and 2018 17th International Conference on Embedded Systems (IEEE), Pune, India, 6–10 January
2018.

\bibitem{das}
J. Das, S. Kevin, R. Srinath, B. Drew,  and S. Bhanja,  ``MRAM PUF: A
novel geometry based magnetic PUF with integrated CMOS'', {\it IEEE
Trans. Nanotechnol.}, {\bf 14}, 436-443 (2015).

\bibitem{zhang}
L. Zhang, X. Fong, C. H. Chang, Z. H. Kong and K. Roy, ``Feasibility
study of emerging non-volatile memory based physical unclonable
functions'', IEEE 6th International Memory Workshop, Taipei, Taiwan, 18–21 May 2014;  pp. 1-4.

\bibitem{marukame}
M. Marukame, T. Tanamoto and Y. Mitani, ``Extracting physically
unclonable function from spin transfer switching characteristics in
magnetic tunnel junctions'', {\it IEEE Trans.  Magn.}, {\bf 50}, pp.1-4 (2014).

\bibitem{vatajelu}
E. I. Vatajelu, G. D. Natale, M. Barbareschi, L. Torres, M. Indaco and
P. Prinetto,  ``STT-MRAM-based PUF architecture exploiting
magnetic tunnel junction fabrication-induced variability'', {\it ACM Journal
on Emerging Technologies in Computing Systems (JETC)}, {\bf 13}, 5 (2016).

\bibitem{rose}
G. S. Rose, N. McDonald, L-K Yan and B. Wysocki,  ``A write-time based
memristive puf for hardware security applications'', in
Proceedings of the 2013 IEEE/ACM International Conference on Computer-Aided Design (ICCAD), San Jose, CA, USA, 18–21
November 2013; pp. 830–833.

\bibitem{winters}
D. Winters, M. A. Abeed, S. Sahoo, A. Barman and S. Bandyopadhyay, ``Reliability of magnetoelastic switching of nonideal nanomagnets with defects: A case study for the viability of straintronic logic and memory'', {\it Phys. Rev. Appl.},  {\bf 12}, 034010 (2019).

\bibitem{mumax}
Available online: https://www.ugent.be/we/solidstatesciences/dynamat/en/mumax (accessed on Sept 20, 2025).



\bibitem{brown}
W. F. Brown Jr., ``Thermal Fluctuations of a Single-Domain Particle'', {\it Phys. Rev.}, {\bf 130}, 1677 (1963).

%\bibitem{book}
%Bandyopadhyay, S. {\it Magnetic Straintronics An Energy-Efficient Hardware Paradigm for Digital and Analog Information Processing}, Synthesis Lectures on Engineering, Science and Technology, Springer Nature, New York, NY, USA, 2022.

%\bibitem{sun1}
%Sun, J. Z. Spin current interaction with a monodomain magnetic body: A model study. {\it Phys. Rev. B}, {\bf 2000}, {\em 62}, 570.

%\bibitem{sun2}
%Sun, J. Z. Current-driven magnetic switching in manganite trilayer junctions, {\it J. Magn. Magn. Mater.}, {\bf 1999}, {\em 202}, 157.

\end{thebibliography}
\end{document}